\documentstyle[aps, psfig, manuscript]{revtex}

\begin{document}

\title{Quantum carpet interferometry for trapped atomic Bose-Einstein condensates}
\author{S. Choi and K. Burnett}
\address{Clarendon Laboratory, 
Department of Physics, University of Oxford, Parks Road, 
\mbox{Oxford OX1 3PU, United Kingdom.}}
\author{O. M. Friesch, B. Kneer, and W. P. Schleich}
\address{Abteilung f\"{u}r Quantenphysik, Universit\"{a}t Ulm, D-89069 Ulm, Germany}

\vspace{6mm}
\maketitle
\begin{abstract}
We propose an ``interferometric'' scheme for Bose-Einstein condensates using near-field diffraction. The scheme is based on the phenomenon of intermode
traces or quantum carpets; 
we show how it may be used in the detection of weak forces.
\end{abstract}


One of the most important aspects of the study of Bose-Einstein Condensates (BEC's) in
dilute trapped alkali gases has been the development of the matter-wave equivalent of optical lasers, the atom laser\cite{atomlaser}. It is essential 
that the development of an atom laser be accompanied by that of useful atom optical 
elements to manipulate the flowing condensates. We have proposed  previously a method to split and
focus a condensate using an external potential\cite{ao}.  In addition to direct focusing and beam splitting, an integral 
part of the field of atom optics is  matter-wave interferometry\cite{AdaSigMly94}. In this 
Brief Report, we propose the use of what is essentially single slit diffraction for matter-waves to perform interferometry with BEC's which should be 
realisable with the current experimental technology. Several theoretical studies of such 
phenomena for ordinary non-condensed wave packets have been carried out 
recently\cite{carpet}, showing in the spacetime probability distribution the so-called ``quantum carpets.'' 
The shapes can be explained by a careful consideration of the solutions of appropriate 
Schr\"{o}dinger's equation.  With the uncondensed atoms, however, at most few atomic wave packets are allowed at a given time in 
order to maintain the quantum coherence required for the spatio-temporal patterns. Clearly this poses a serious experimental challenge; for instance, an exceedingly good vacuum system would be needed to maintain the carpet for a reasonable duration of time. An immediate 
advantage of BEC is that the patterns can be formed with $N$ atoms, which allows  possible experimental observation of the atomic quantum carpets, and its application as a matter-wave interferometer. 

In order to construct a quantum carpet, we begin with a Bose-Einstein condensate close to temperature $T=0$, stored 
in a one dimensional harmonic trap of frequency $\omega$. (The minimum of
the trap is set to be at $x=0$.) At time $t=0$ we suddenly switch off the trap and release the
condensate in a ``harder'' 
potential of the form  ${\bf r}^{\xi}$  where $\xi \gg 2$, with the walls symmetrically located at
$x=\pm L$.   
For simplicity, we propose to use 
an inverted super-Gaussian, $V_{\rm box}({\bf r})$ as our potential:
\begin{equation}
V_{\rm box}({\bf r}) = \frac{L^2}{2} \left \{1- \exp \left [ - \left (\frac{\bf r}{\sigma} \right )^{m} \right ] \right \},
\end{equation}  
where $L$ is defined to be half of the total transverse length of the potential in harmonic oscillator units of length, and $\sigma = 0.9L$. With $m = 42$, the 
potential closely approximates a box of length $2L$. Experimentally, the potential is expected to be well approximated by using parallel sheets of optical fields. It is clear that $V_{\rm box}$ is equivalent in the optical context to imposing a slit, which simply provides a set of boundary conditions for Maxwell's equations.  We note that the group of Ertmer has already implemented coherent reflection of BEC from an atom optical mirror and also investigated BEC in an atom-optical wave guide, which are very closely related to the experimental technology required to observe a quantum carpet\cite{Ertmer}.    

We describe the dynamical evolution of Bose-Einstein condensates using the Gross-Pitaevskii 
Equation (GPE), which provides an excellent description of BEC at temperatures $T \ll T_c$ 
where $T_c$ is the critical temperature of phase transition as confirmed by various experiments
 to date. This equation has the form:
\begin{equation}
i \hbar \frac{\partial \psi}{\partial t}  = \left [ -\frac{\hbar^2}{2M}\nabla^{2}_{\bf r} +  
V({\bf r}) + C |\psi|^2 \right ] \psi, \label{GP}
\end{equation}
with $C = NU_{0} = \frac{4\pi\hbar^{2} N a}{M}$,
where $N$ is the number of particles, $a$ the interatomic $s$-wave scattering length 
and $M$ the mass of a single particle. $V({\bf r})$ is the trapping potential which may be switched off and replaced by $V_{\rm box}({\bf r})$. 
The prepared condensate wave packet of parabolic shape is evolved according 
to the one dimensional version of Gross-Pitaevskii equation:
\begin{equation}
i \hbar \frac{\partial \psi}{\partial t}  = \left [ -\frac{\hbar^2}{2M}
\frac{\partial^{2}}{\partial x^2} + V_{\rm box}({\bf r}) \kappa |\psi|^2 \right ] \psi.
\end{equation}
Here the size, $\kappa \equiv C / (\pi a_{\rm HO}^2)$ of the nonlinearity
is a reduction from the three- to the one-dimensional case. 
Such 1D simulation does indeed have a relevance in reality. For example, a 3D trap with a box potential in one direction, along with tightly confining 
harmonic potentials in the remaining two directions may be used. Assuming
weak directional coupling in the condensate motion, this would then produce  interference in one direction only, giving results effectively for a  1D
carpet.
	
\begin{figure}[t] \begin{center}
\centerline{\psfig{height=7cm,file=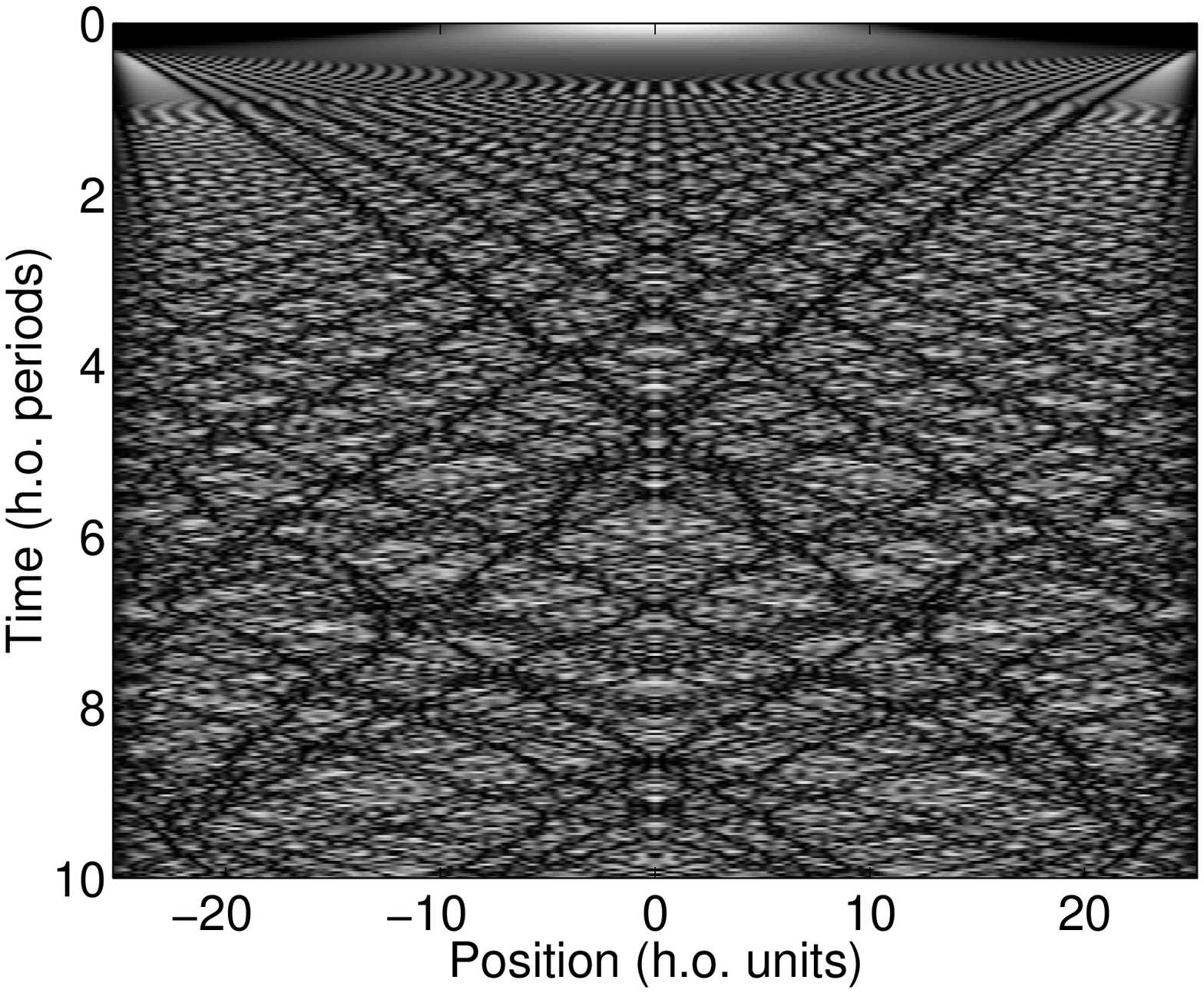}}
\end{center}
\caption{{\protect \footnotesize  Quantum carpet with BEC, represented as spacetime probability density. The time and position are given in units of harmonic oscillator (h.o.) period and length respectively.}}  \label{carpetFig1}
\end{figure}

As a result of switching to $V_{\rm box}$, the kinetic energy term in GPE dominates and the condensate  wave packet rapidly spreads as function of time, and reflects at the walls acting as atomic mirrors. It then interferes with itself with varying degree over time producing complex intensity profiles 
or ``fringe patterns.'' Shortly afterwards when these fringes are viewed in the time domain,
regular dark canals of low density are formed and we observe a remarkable emergence of spatio-temporal patterns\cite{footnote}.  The pattern for $L=25 a_{\rm HO}$, $T=0$ and evolution time of $0$ to $10\tau _{\rm HO}$ where $a_{\rm HO}$ is one harmonic oscillator length and
$\tau _{\rm HO}$ is one period of the harmonic oscillator is shown in Fig.~\ref{carpetFig1}. The total duration of simulation is therefore approximately 100 ms for a trap with frequency 
$\omega = 2\pi \times 100 {\rm Hz}$. 
For this simulation, we 
used a nonlinearity $\kappa = 500 \, \hbar \omega \pi a_{\rm HO}$ which corresponds to approximately
20000 Rubidium atoms. 
 Qualitatively
similar canals and ridges occur when we propagate the same wave packet according to
the GPE with $\kappa = 0$ (i.e. linear, uncondensed case). However three fundamental
differences occur in the carpet with the nonlinearity turned on: ({\it i}) The initial spreading is much faster; this can be related to the time dependent excitation of higher 
momenta due to interactions. ({\it ii}) The canals and ridges develop
a substructure along the straight lines predicted by the linear propagation. 
({\it iii}) We note complicated avoided crossings when two canals cross each other.  The most important difference is that even for much longer integration times the fractional revivals of the 
wave function do not appear as seen in the linear case.  We note that the sizes and the resolution of the observed patterns 
depend strongly on factors such as the strength of nonlinearity $\kappa$ and the length $2L$ of the box. 
In particular, the pattern becomes more distinct and the canals turn to ridges and vice versa with fewer 
number of atoms or smaller $\kappa$. 
From an experimental point of view, we note that sufficient 
temporal resolution to observe the carpet structure exists with, for instance, the nondestructive imaging procedure which has the shortest 
probe pulse duration of order 1 ms\cite{MITimaging}.  In
particular, with a set of repeated measurements made at various different reference times one may expect to improve the resolution further. 

We note that the quantum carpets woven with uncondensed atomic wave packets have been extensively analysed using various methods. They are found to exhibit regular spatio-temporal patterns consisting of straight lines of varying gradients, representing harmonics of a fundamental wave velocity $v_{f} = \omega_{f}/k_{f} = \hbar \pi / (2mL)$, where $\omega_f = \hbar \pi^{2}/(2ML^{2})$ and $k_f = \pi /L$ denote the fundamental frequency and wave number in a box potential of length $L$. The probability density results from the interferences between degenerate eigenmodes of the system. The straight lines in the carpet formed from uncondensed wave packets represent wave velocities in integer multiples of $v_{f}$, $v_{ln} = (l \pm n)v_{f}$ where $l$ and $n$ are integers, $l,n = 0,1, \ldots$.  One expects similar analysis for the quantum carpets with BEC, and indeed as $\kappa \rightarrow 0$ this is precisely what happens with the BEC quantum carpets. 

However, for large values of $\kappa$, the self-interaction present in BEC  considerably complicates the analysis of the carpet, leaving numerical analysis as essentially the only viable option. However, it is found that one can gain useful insight into the existing structures by combining two facts about BEC.  One is that, depending on the sign of the $s$-wave scattering length, condensates are able to support bright, grey and dark solitons which are the unique features of nonlinear wave equations such as the GPE. In fact, under GPE, any solution which is not an eigenstate of the given external potential has a tendency to form solitons as $t \rightarrow \infty$. In particular, a study of grey soliton formation in a condensate resulting from collisions have been carried out by Scott {\it et al.}\cite{Scott} Grey solitons, which are the localised reduction in the ambient density, or the ``density notch'' solitons, are found in inhomogeneous condensates with positive $s$-wave scattering lengths, and they are found to play an important role in governing the bulk motion of the condensate. They tend to store energy in a slow moving structure and act as a throttle to retard the movement of the condensate, possibly playing an important role in dissipating superfluid flow\cite{Scott}. Their speed of propagation $v$ is controlled by the difference in phase $\delta$ between the parts of the condensate separated by the density dip, in a form reminiscent of Josephson effect\cite{Reinhardt}: $v = c_{\rm B} \/ \sin(\delta/2)$ where $c_{\rm B} = \sqrt{4 \pi \hbar^{2} a \rho/m^{2}}$ is the Bogoliubov speed of sound  with $\rho$ being the condensate density. A corollary is that the speed of grey solitons provides a sensitive measure of the phase difference between adjoining regions of the condensate.

The second fact comes from the work on soliton dynamics in BEC. Kosevish\cite{Kosevich} and also Reinhardt and Clark\cite{Reinhardt} have shown that, in the case where $V_{\rm trap}$ varies slowly on the scale of the condensate correlation length, the grey solitons behaves as a ``classical'' particle of positive mass. The position of the center of the soliton $x_{c}(t)$ obeys the Newtonian equation of motion:    
\begin{equation}
\ddot{x}_{c}(t) = - \frac{\partial V_{\rm trap}}{\partial x_{c}}.
\end{equation}
This has been demonstrated using $V_{\rm trap}$ of harmonic form, which clearly shows the soliton in a simple harmonic motion. The main point to note is that the solitons exhibit particle-like behaviour, demonstrating, quite remarkably, the possibility of applying classical Newtonian mechanics in the description of a purely quantum system where wave-mechanics is supposed to dominate its behaviour. 

Ignoring the more fine lines (smaller solitons) one can see immediately about 30 main dark straight lines in Fig.~\ref{carpetFig1} criss-crossing each other ($V_{\rm trap} = 0$ for $-L < x < L$). They represent grey solitons bounced off the side walls, each travelling at some constant speed, determined by the amount of phase slip the soliton gains as a result of interference that occurs when the condensates reflect from the walls. 
The different velocities may be understood from the fact that upon reflecting and interfering, the condensate gains phase gradient which is step-like, yet with different gradient per step as shown in Scott {\it et al.}\cite{Scott}. 
One can see that as soon as the condensate meets the wall at time $\sim 1 \pi/\omega$, it creates (or ``launches'') approximately 15 major grey solitons travelling at different constant speeds from either side. Faster solitons move further away from the walls within the same time duration than the slower ones, and reach the opposite wall earlier. The largest velocity is approximately that of the Bogoliubov speed in average atomic density.   Once reaching the opposite wall the solitons rebound but simultaneously generate even more solitons of smaller sizes. It is also noted that the smaller grey solitons represented by the finer lines on the carpet continually collide and interact with other solitons to eventually alter the paths, sizes and shapes of the major grey solitons quite significantly from the simpler Newtonian picture. However, initially at least i.e. up to time of $\sim 15 \pi/\omega$, the main grey solitons can clearly be observed to obey the Newtonian mechanics. The avoided crossings are indicative of the repulsive interaction.

\begin{figure}[t] \begin{center}
\centerline{\psfig{height=7cm,file=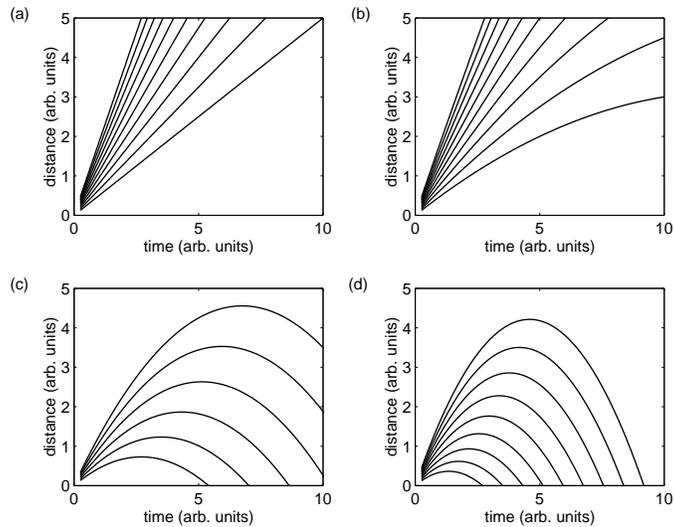}}
\end{center}
\caption{{\protect \footnotesize  Newtonian trajectory of classical particles under the potentials of the form (a) ${\cal V}(x) = 0$ (b) ${\cal V}(x) = 0.01x$ (c) ${\cal V}(x) = 0.05x$ (d) ${\cal V}(x) = 0.1x$}, where $x$ is along the vertical axis. The different lines correspond to different initial velocities.}  \label{newton}
\end{figure}

\begin{figure}[t] \begin{center}
\centerline{\psfig{height=7cm,file=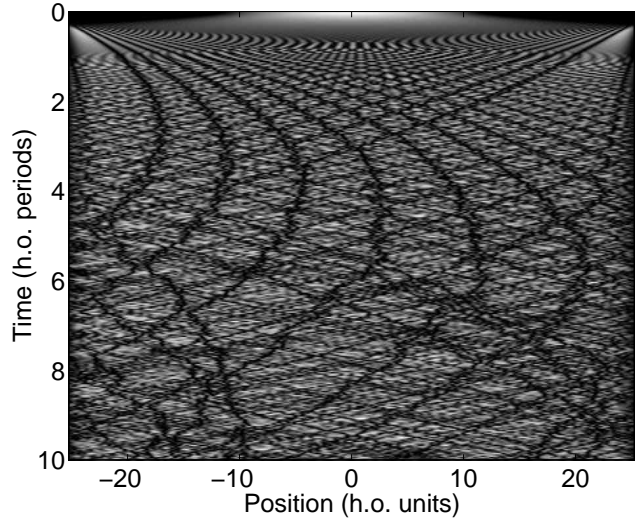}}
\end{center}
\caption{{\protect \footnotesize  Quantum carpet with BEC, in an externally imposed potential of the form ${\cal V}(x) = 0.05x$} in harmonic oscillator (h.o.) units.}  \label{carpetFig2}
\end{figure}

\begin{figure}[t] \begin{center}
\centerline{\psfig{height=7cm,file=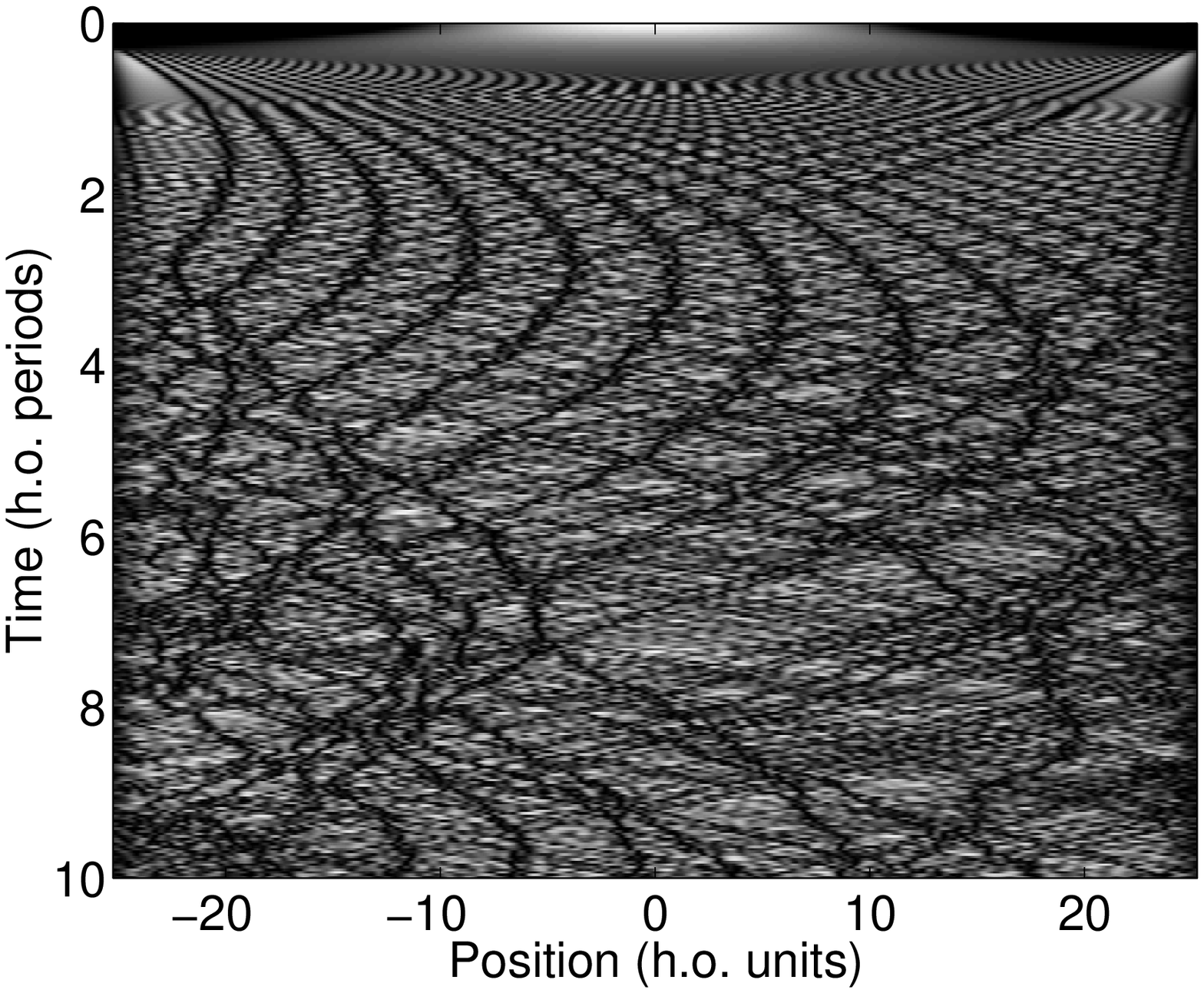}}
\end{center}
\caption{{\protect \footnotesize  Quantum carpet with BEC, in an external potential of the form ${\cal V}(x) = 0.1x$.}}  \label{carpetFig3}
\end{figure}

We now illustrate the effect of adding a linear variation ${\cal V}$ to the  inverted super-Gaussian potential of the quantum carpet. Such a potential models the effect of an external force, and subsequent phase shift across the quantum system.  Expected Newtonian trajectories for particles travelling at different but evenly distributed initial speeds under several different constant accelerations are given in Fig.~\ref{newton}. Fig.~\ref{carpetFig2} shows the BEC quantum carpet where ${\cal V} = 0.05x$ in scaled units and  Fig.~\ref{carpetFig3} the case where ${\cal V} = 0.1x$ in scaled units. In real units (in the case the trap frequency is $2\pi \times 100$ Hz and using Rubidium atoms) these correspond to $0.003g$ and $0.006g$ respectively where $g$ is the acceleration due to gravity. It is quite clear that the canals
show {\em exactly} what one expects from standard kinematics.
This result shows that one can deduce the nature and magnitude of the potential (or equivalently interferometric phase shifts) from observing the motion of the grey solitons (or the interference fringes). This allows not only the ability to detect standard time-independent linear potential but, in principle, also gives us the ability to deduce time-varying potential of different shapes, as long as  the corresponding Newtonian motion can be identified.  

The limitation of any interferometric method would be the intrinsic phase diffusion within the condensate which may eventually destroy the interference patterns. In real experiments, the phase diffusion rate is expected not to differ too greatly from that of the original MIT interference experiment\cite{MITinterference}, which reported a decoherence time of the order of 100 ms.  Assuming that the three-body recombination rate is more or less similar for different atomic species, the diffusion time scale is clearly much longer than the integration times we have presented here, reflecting the surprising robustness of the BEC interference fringes against decoherence. We note also that recently Burger {\it et al.} has observed the evolution of dark solitons in BEC, created using a phase imprinting method\cite{Burger}. Obviously their experiment is not a quantum carpet phenomenon; however it is clear that the force measurements discussed in this paper may directly be implemented in their experiment. 

In summary, we have studied how one may use the quantum carpets with BEC as an interferometer. The spatio-temporal patterns could be explained using the idea of soliton formation, along with the classical equation of motion for the solitons. 
This is a very different kind of interferometer than the standard matter wave interferometer such as the Mach-Zehnder, and is of interest from a number of different aspects.  

We acknowledge the support from the European Union under the TMR Network Programme. SC wishes to thank Prof. R. J. Ballagh and J. Ruostekoski for valuable discussions. BK acknowledges gratefully the support of the Deutsche Forschungsgemeinschaft (Forschergruppe Quantengase).
OMF and BK acknowledge support of the DAAD.

\vspace{10mm}

\begin{references}
\bibitem{footnote} We note that, as to be expected, the exact shape of the carpet pattern is dependent on the potential $V_{\rm box}$ used.
\end{references}
\end{document}